\documentclass[lettersize,journal]{IEEEtran}
\usepackage{amsmath,amsfonts}
\usepackage{algorithmic}
\usepackage{array}
\usepackage[caption=false,font=normalsize,labelfont=sf,textfont=sf]{subfig}
\usepackage{textcomp}
\usepackage{stfloats}
\usepackage{url}
\usepackage{verbatim}
\usepackage{graphicx}
\usepackage{caption}
\usepackage{subcaption}
\usepackage{xcolor}
\usepackage[x11names]{xcolor}
\usepackage{hyperref}
\def\BibTeX{{\rm B\kern-.05em{\sc i\kern-.025em b}\kern-.08em
    T\kern-.1667em\lower.7ex\hbox{E}\kern-.125emX}}
\usepackage{balance}

\usepackage[
    style=ieee,
    url=false,
    isbn=false,
    ]{biblatex}
\begin{document}
\title{The SWEET project: probing sugar crystals for direct dark matter searches}
\author{
\IEEEauthorblockN{
A. Bento\IEEEauthorrefmark{1}\IEEEauthorrefmark{2},
F. Casadei\IEEEauthorrefmark{1},
E. Cipelli\IEEEauthorrefmark{1},
S. Di Lorenzo\IEEEauthorrefmark{1},
F. Dominsky\IEEEauthorrefmark{1}\IEEEauthorrefmark{4},
P. V. Guillaumon\IEEEauthorrefmark{1}\IEEEauthorrefmark{3}\IEEEauthorrefmark{5},
D. Hauff\IEEEauthorrefmark{1},\\
A. Langenkämper\IEEEauthorrefmark{1},
M. Mancuso\IEEEauthorrefmark{1},
B. Mauri\IEEEauthorrefmark{1}\IEEEauthorrefmark{6},
C. Moore\IEEEauthorrefmark{1}\IEEEauthorrefmark{7},
F. Petricca\IEEEauthorrefmark{1},
F. Pröbst\IEEEauthorrefmark{1},
M. Zanirato\IEEEauthorrefmark{1}
}
\IEEEauthorblockA{
\IEEEauthorrefmark{1}Max-Planck-Institut für Physik, D-85748 Garching, Germany\\
\IEEEauthorrefmark{2}LIBPhys-UC, Departamento de Fisica, Universidade de Coimbra, P3004 516 Coimbra, Portugal\\
\IEEEauthorrefmark{3}Instituto de Física da Universidade de São Paulo, São Paulo 05508-090, Brazil\\
Email: \IEEEauthorrefmark{4}dominsky@mpp.mpg.de,
\IEEEauthorrefmark{5}pedro.guillaumon@mpp.mpg.de,
\IEEEauthorrefmark{6}bmauri@mpp.mpg.de
\IEEEauthorrefmark{7}moore@mpp.mpg.de
}
}

\markboth{Journal of \LaTeX\ Class Files,~Vol.~18, No.~9, September~2020}%
{How to Use the IEEEtran \LaTeX \ Templates}

\maketitle

\begin{abstract}
Several experiments are currently competing to directly detect sub-GeV/c$^2$ dark matter through elastic scattering off various target nuclei.  Hydrogen-rich materials, such as organic compounds, are promising candidate targets for such experiments due to favourable kinematics.

In this paper, we present for the first time results obtained with a sugar-based cryogenic particle detector. A sucrose ($\mathbf{C_{12}H_{22}O_{11}}$) monocrystal was instrumented with a neutron transmutation doped germanium thermistor to detect phonons from particle interactions, and was operated in the vicinity of a light detector.  Particle interactions in the sugar were observed with associated scintillation light, indicating a possible channel for particle discrimination.

\end{abstract}

\footnotetext{This work has been submitted to the IEEE for possible publication. 
Copyright may be transferred without notice, after which this version may no longer be accessible.}


\section{Introduction}
The identity of dark matter (DM) remains an open question in modern physics. Weakly interacting massive particles (WIMPs) with a mass $\mathcal{O}$(GeV-TeV/c$^2$) have been one of the most studied DM candidates \cite{ParticleDataGroup:2024cfk,cirelli2024darkmatter}, however compelling sub-GeV/c$^2$ models also exists, so-called light dark matter, which represents a well-motivated and largely unexplored mass range beyond the traditional WIMP paradigm. Theoretical models involving light mediators or hidden sectors naturally predict such candidates, which remain consistent with cosmological constraints \cite{Blennow_2017,Feng_2008,BOEHM2004219,PhysRevD.91.023512,
cheung2011origins}. Several collaborations, including CRESST \cite{cresst_SP}, SuperCDMS \cite{supercdmscollaboration_SearchLowmassDark_2023}, TESSERACT \cite{chang2025first}, DELight\cite{von2023delight}, and BULLKID \cite{colantoni2020bullkid}, are developing next-generation detectors with the goal of accessing lower dark matter masses with unprecedented sensitivity.

To enhance sensitivity to sub-GeV/c$^2$ DM masses, lighter target nuclei are required, as they enable more efficient momentum transfer in low-mass DM interactions. Since hydrogen is the lightest element, it would be the ideal target for accessing the lowest masses of DM \cite{wang_LightDarkMatter_2022a}.  Sucrose, an organic crystal with the molecular formula $\mathrm{C_{12}H_{22}O_{11}}$, offers an attractive combination of hydrogen, carbon, and oxygen nuclei, allowing for sensitivity across a broader range of dark matter masses. These features, in addition to its low cost, position sucrose as a compelling target for future low-temperature detectors exploring the sub-GeV/$c^2$ dark matter parameter space.

The goal of the SWEET project is to investigate the suitability of sugar crystals as potential dark matter detectors. In this paper, we report the first results of a sucrose monocrystal operated as a particle detector at millikelvin temperatures.

\section{Dark matter searches with sugar}

\begin{figure}[h]
    \centering
    \includegraphics[width=1\linewidth]{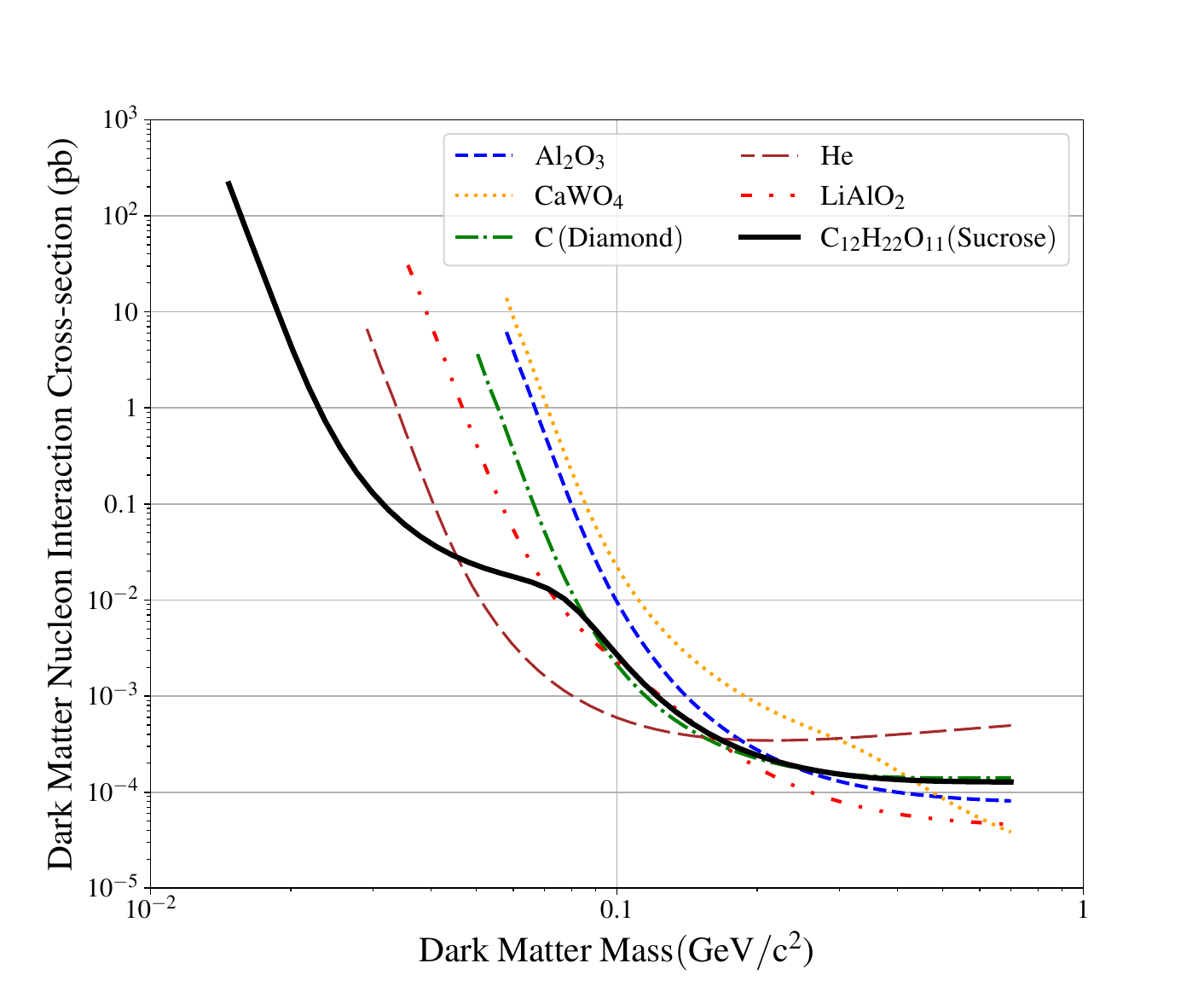}
    \caption{Projected exclusion limits on the elastic, spin-independent DM-nucleon interaction cross section for sapphire (Al$_2$O$_3$), calcium tungstate (CaWO$_4$), helium (He), lithium aluminate (LiAlO$_2$), carbon (diamond), and sucrose ($\mathrm{C_{12}H_{22}O_{11}}$), under the assumptions outlined in the text.
    }
    \label{fig:sugar-limit}
\end{figure}

To evaluate its suitability for dark matter direct detection, the projected exclusion limits on the spin-independent elastic DM-nucleon scattering for sucrose are compared with those of other materials previously deployed for R\&D: diamond (C) \cite{Diamond}, sapphire (Al$_2$O$_3$) \cite{cresst_SP}, calcium tungstate (CaWO$_4$) \cite{CaWO4_CRESST}, and lithium aluminate (LiAlO$_2$) \cite{Lithium_CRESST}, as well as helium due to its low mass and recent interest for sub-GeV/c$^2$ dark matter searches \cite{von2023delight,herald}.  These limits are produced assuming zero background for an exposure of 1\,kg$\cdot$day, an energy threshold of 5\,eV, and the standard halo model with $\rho_{DM} = 0.3$\,GeV/(c$^2\cdot$cm$^3$), $v_{esc} = 544$\,km/s, $v_{Earth} = 232$\,km/s, and characteristic WIMP velocity $v_0 = 220$\,km/s.  The results, shown in Figure~\ref{fig:sugar-limit}, indicate that among the targets considered, sugar is able to access the lowest masses of WIMP-like DM, provided that the phonons produced by particle interactions are able to be efficiently collected and backgrounds can be suppressed.

In addition to the conventional spin-independent WIMP-nucleon interaction explored above, the presence of unpaired protons in organic crystals makes them an interesting candidate material for the exploration of spin-dependent WIMP-proton interactions \cite{hydrogen_1}.

\section{Assembly of the sugar-based detector}
\noindent The aim of this study is to investigate whether sugar crystals exhibit detectable particle signals and whether they emit scintillation light. To this end, a dedicated detector module with phonon and light readout channels was designed and assembled.

\begin{figure}[!h]
\centering

\begin{minipage}[b]{0.12\textwidth}
  \centering
  \includegraphics[width=\linewidth]{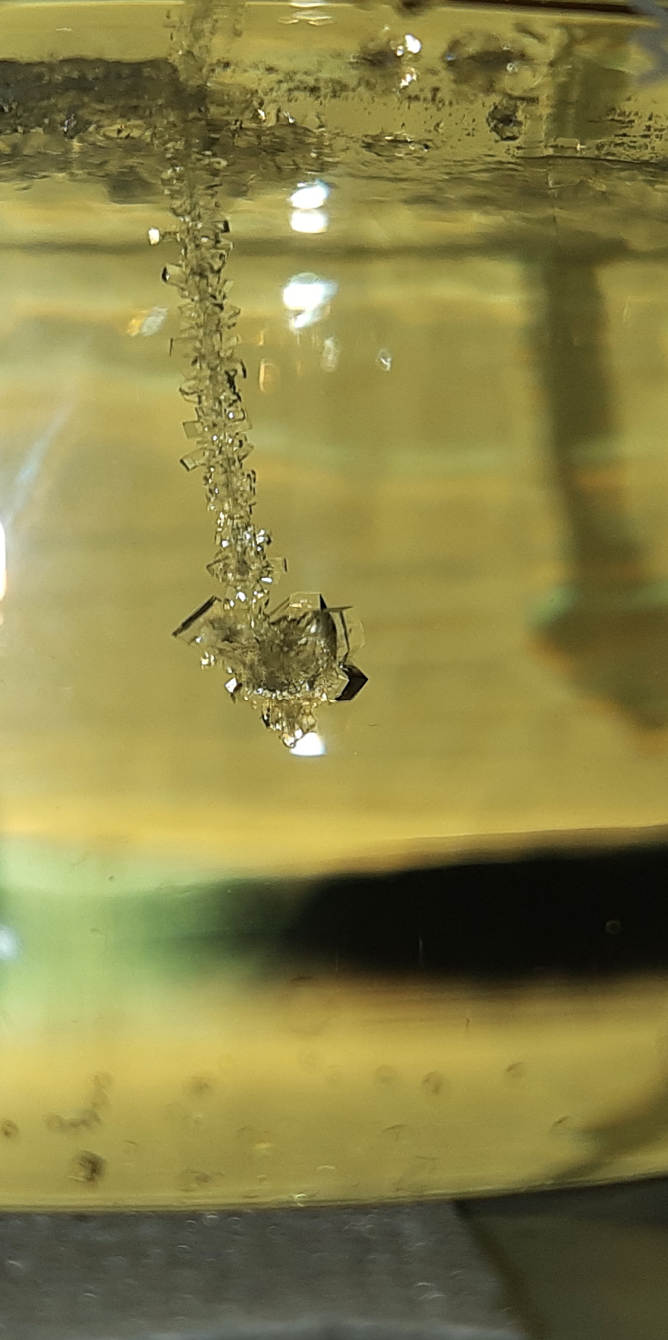}
  \small (a)
\end{minipage}
\hfill
\begin{minipage}[b]{0.15\textwidth}
  \centering
  \includegraphics[width=\linewidth]{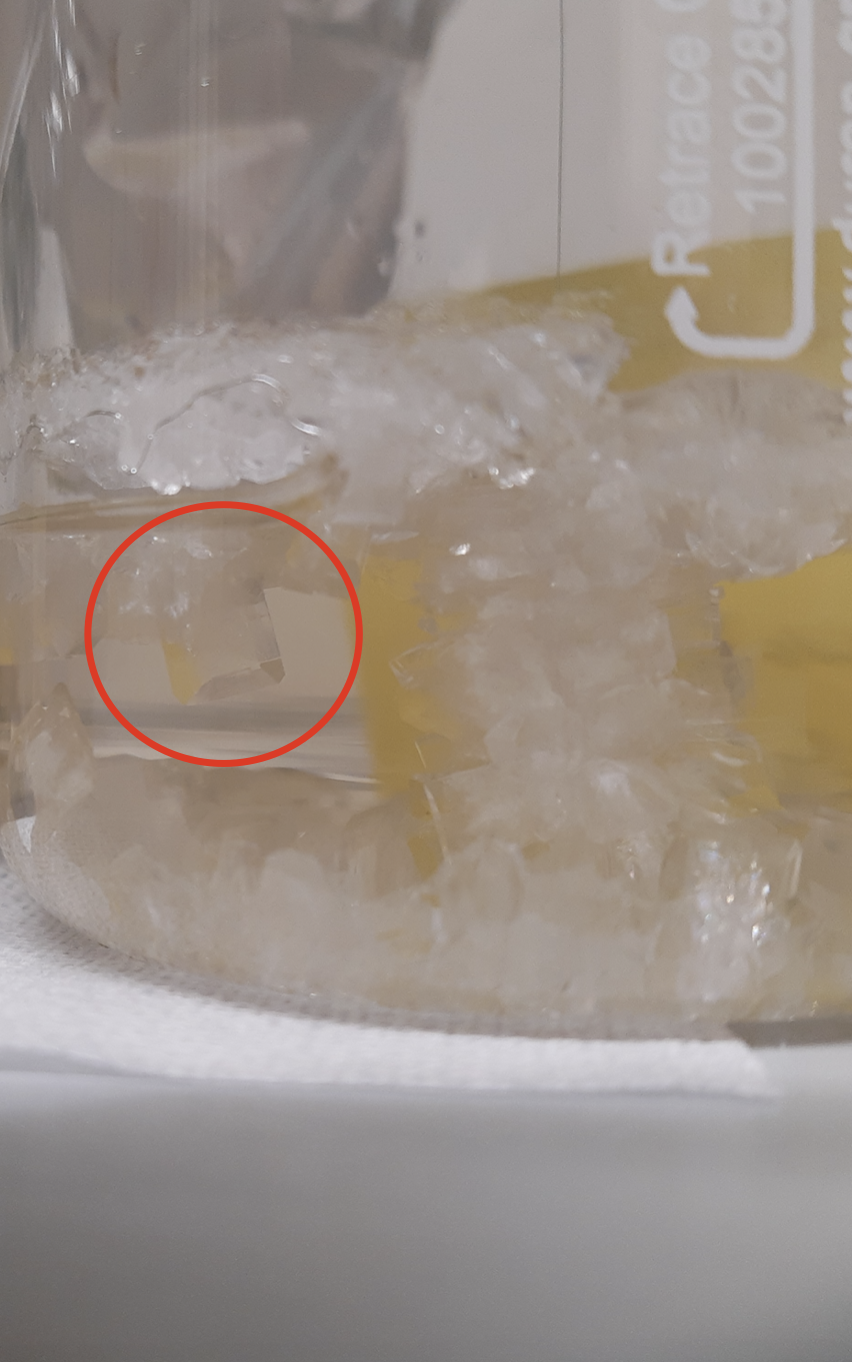}
  \small (b)
\end{minipage}
\hfill
\begin{minipage}[b]{0.119\textwidth}
  \centering
  \includegraphics[width=\linewidth]{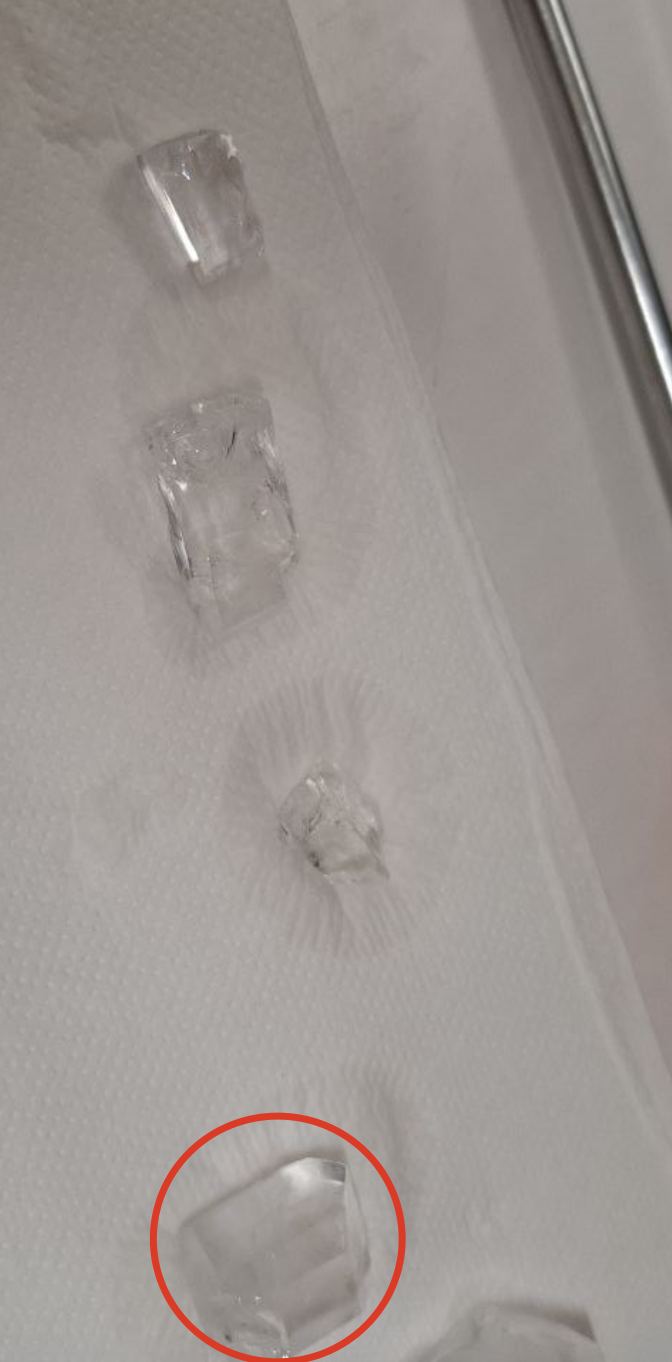}
  \small (c)
\end{minipage}

\caption{(a) Initial sugar crystallization on a suspended nylon wire immersed in the supersaturated solution; (b) Sugar crystals formed after several weeks of slow recrystallization. An example of a monocrystalline crystal is indicated in the circle; (c) Sugar crystals collected from the supersaturated solution. The crystal highlighted in the circle was selected for the prototype detector described in this work.}
\label{fig:sugar-growth}
\end{figure}

To produce monocrystalline sugar samples, a crystal growth technique was followed based on the principle of slow recrystallization from a supersaturated sucrose solution. A 3:1 ratio by mass of commercially available sugar to deionized water was heated until fully dissolved. The solution was then gradually cooled in a sealed container to avoid surface crystallization. The initial crystallization was induced on suspended nylon wires (Figure \ref{fig:sugar-growth}), and over several weeks, monocrystalline structures of sufficient size were collected and their surfaces polished. For the initial test, a single sugar crystal was selected with an approximately regular shape and a mass of 0.96\,g.

\begin{figure}[!h]
\centering

\begin{minipage}[b]{0.223\textwidth}
  \centering
  \includegraphics[width=\linewidth]{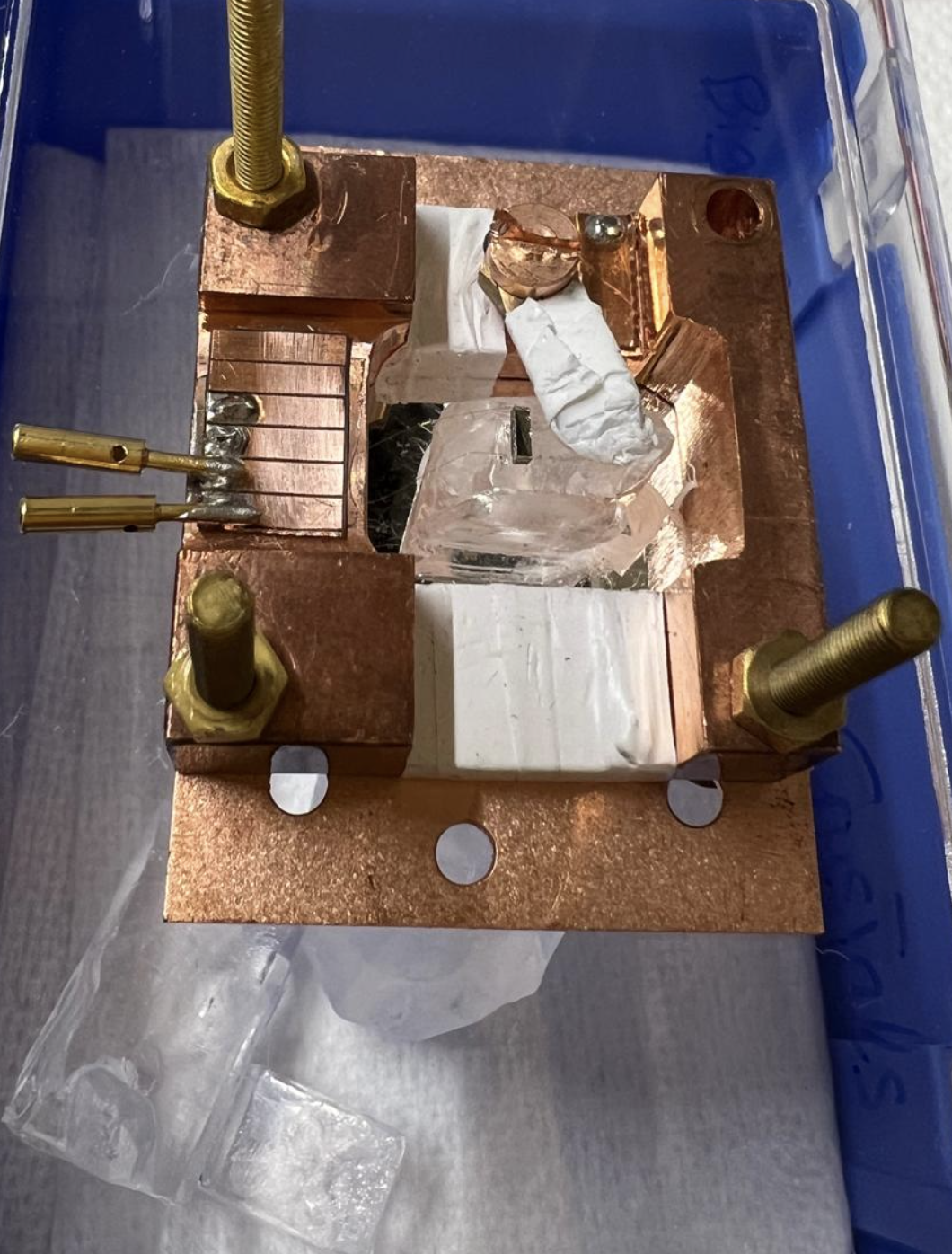}
  \small (a)
\end{minipage}
\hfill
\begin{minipage}[b]{0.2\textwidth}
  \centering
  \includegraphics[width=\linewidth]{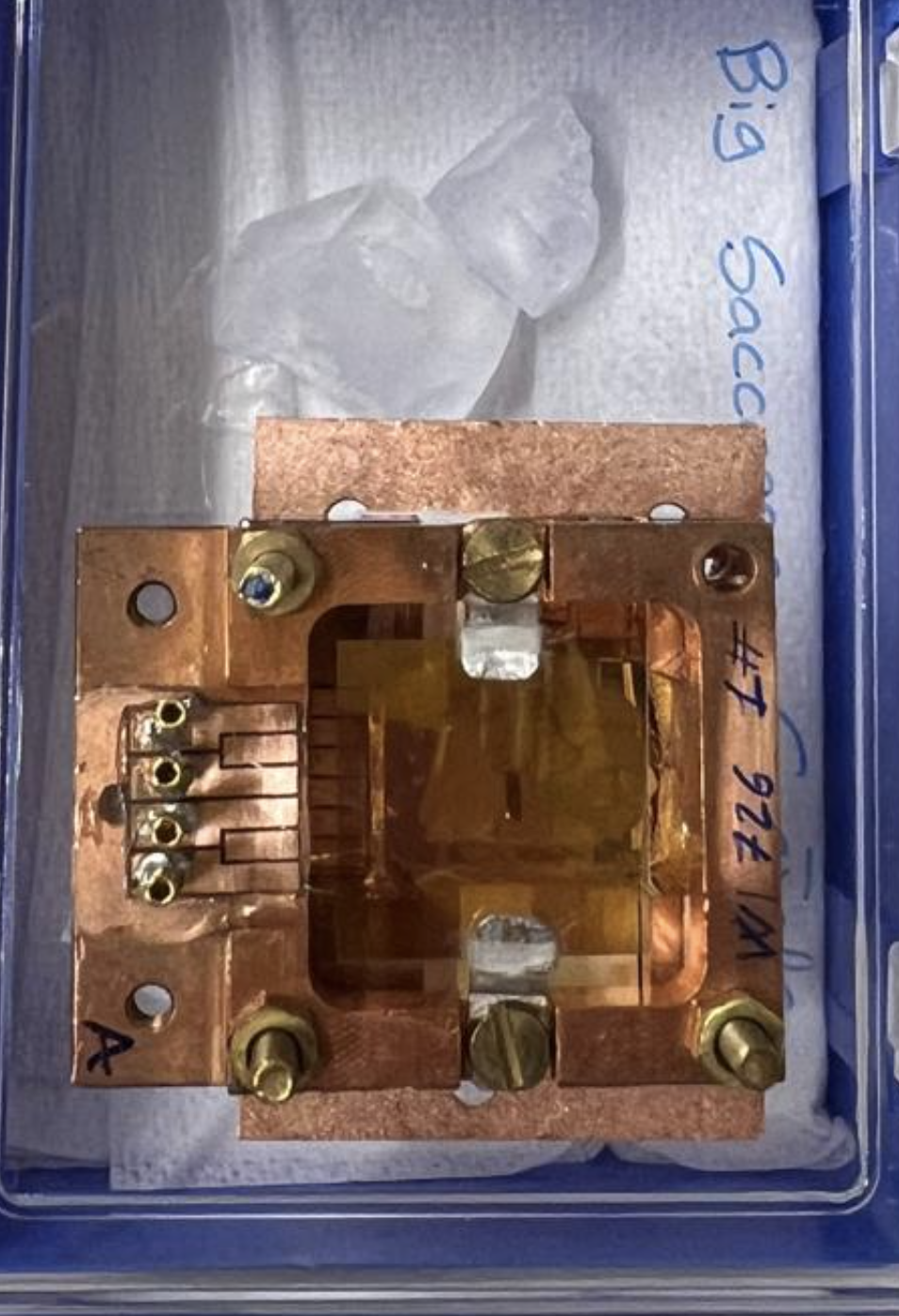}
  \small (b)
\end{minipage}

\caption{(a) Sugar crystal instrumented with an NTD thermistor and mounted in its copper holder. Larger sugar crystals, produced using the same procedure, are visible in the box below the detector; (b) Light detector module mounted above the sugar detector.}
\label{fig:sugar-detector}
\end{figure}

The chosen crystal was instrumented with a neutron transmutation doped (NTD) germanium thermistor \cite{larrabee2013neutron} (1×1×3 mm$^3$), glued using three small dots of epoxy (Gößl+Pfaff GP 12) applied through a Mylar mask to ensure precise placement and prevent merging of the glue spots (Figure \ref{fig:sugar-detector}a).

The sugar crystal was mounted in a copper holder resting on thin polytetrafluoroethylene (PTFE) supports. These elements provided mechanical stability during thermal contraction and served as a weak thermal link between the crystal and the heat sink. The crystal was secured to the holder with a clamp wrapped in PTFE tape to protect the polished surface. Electrical contacts to the NTD thermistor were made using single 25\,$\mu$m diameter gold wires, bonded to copper-kapton-copper pads.

To investigate potential scintillation, the sugar crystal was accompanied by a light detector (Figure \ref{fig:sugar-detector}b). It consisted of a 20×20×0.4\,mm$^3$ silicon-on-sapphire crystal \cite{cresst_SP} with a transition edge sensor (TES) microfabricated directly onto the surface \cite{angloher2024doubletes,angloher2024detector}. A reflective foil was placed below the sugar crystal and above the light detector to enhance light collection efficiency. A continuous data stream was recorded simultaneously with that of the sugar crystal to capture coincident events.

The complete detector assembly was installed and tested at cryogenic temperatures in an Oxford Instruments dilution refrigerator at the Max Planck Institute for Physics in Garching bei München (Germany) with a base temperature below 7\,mK. The data were recorded as a continuous stream with 50\,kHz sampling frequency, using a 16 bit digitizer from National Instruments (NI USB-6218 BNC).

\section{Preliminary observations}
Approximately 19 hours of continuous stream data were acquired during the measurement campaign. Individual optimum filters \cite{OFilter} were implemented for both the phonon and light channels using an averaged pulse template derived from representative signal events and noise-power spectra. The phonon channel was used as the primary trigger for the data acquisition, with a threshold set at 20\,mV. Given a baseline resolution of 1.4\,mV on the sugar detector, this conservative threshold ensured that the impact of noise on the results is negligible.

\begin{figure}[h]
    \centering
    \includegraphics[width=1\linewidth]{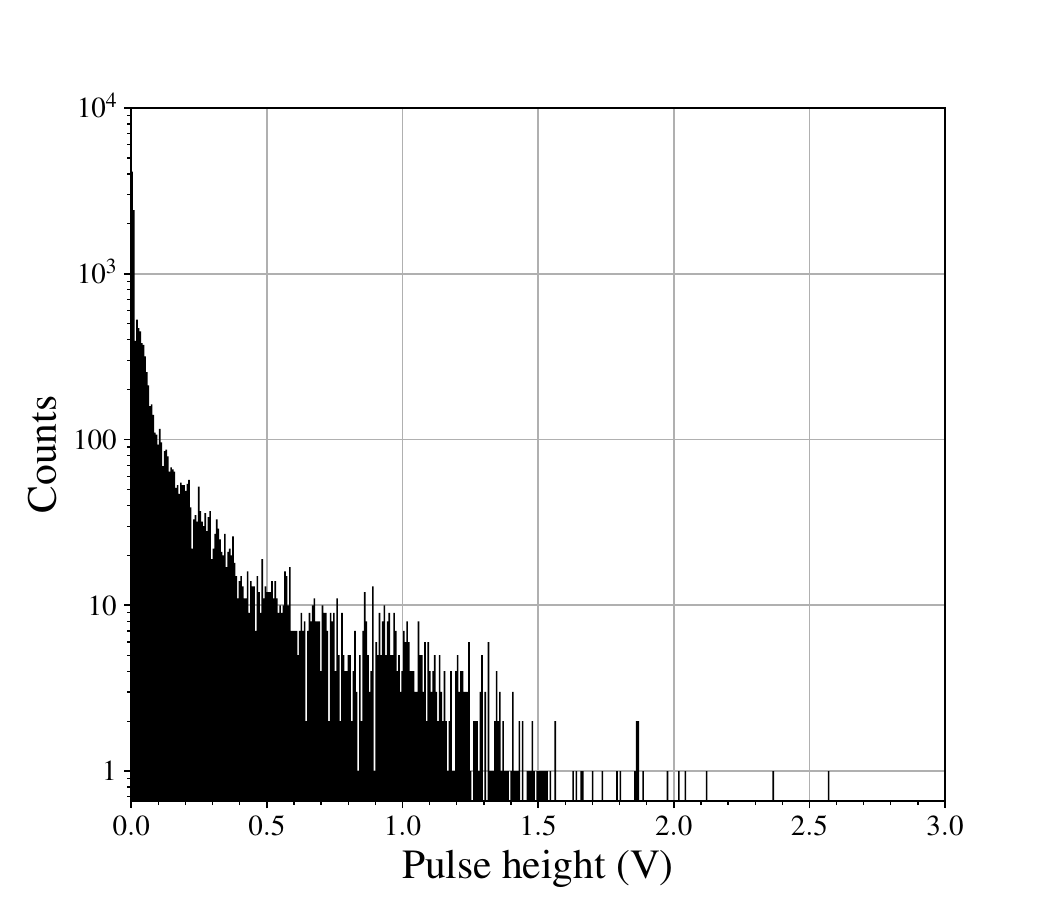}
    \caption{Distribution of pulse amplitudes observed in the sugar crystal. The amplitudes were obtained by applying an optimum filter to all triggered events.}
    \label{fig:sugar-spec}
\end{figure}

After triggering, quality cuts were applied based on noise and pulse parameters. The resulting filtered pulse spectrum for the sugar detector is shown in Figure \ref{fig:sugar-spec}. While the spectrum does not exhibit distinct features suitable for an energy calibration, particle interactions within the crystal were successfully detected, suggesting that sugar remains a promising candidate material worthy of further investigation. For future campaigns, the setup will be improved with a dedicated X-ray source to provide a well-defined calibration signal.

\begin{figure}[h!]
    \centering
    \includegraphics[width=1\linewidth]{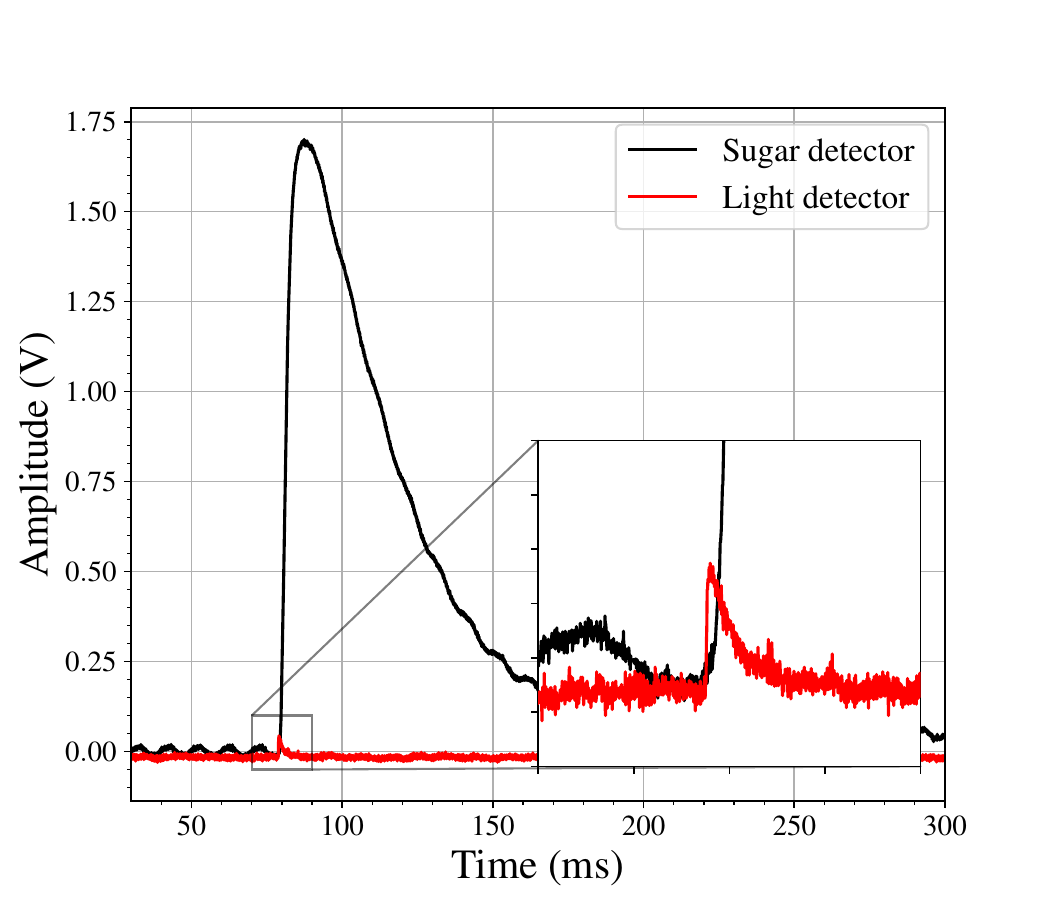}
    \caption{Example of coincident events detected simultaneously in the phonon (black) and the light (red) channels.}
    \label{fig:sugar-pulse}
\end{figure}

\begin{figure}[h!]
    \centering
    \includegraphics[width=1\linewidth]{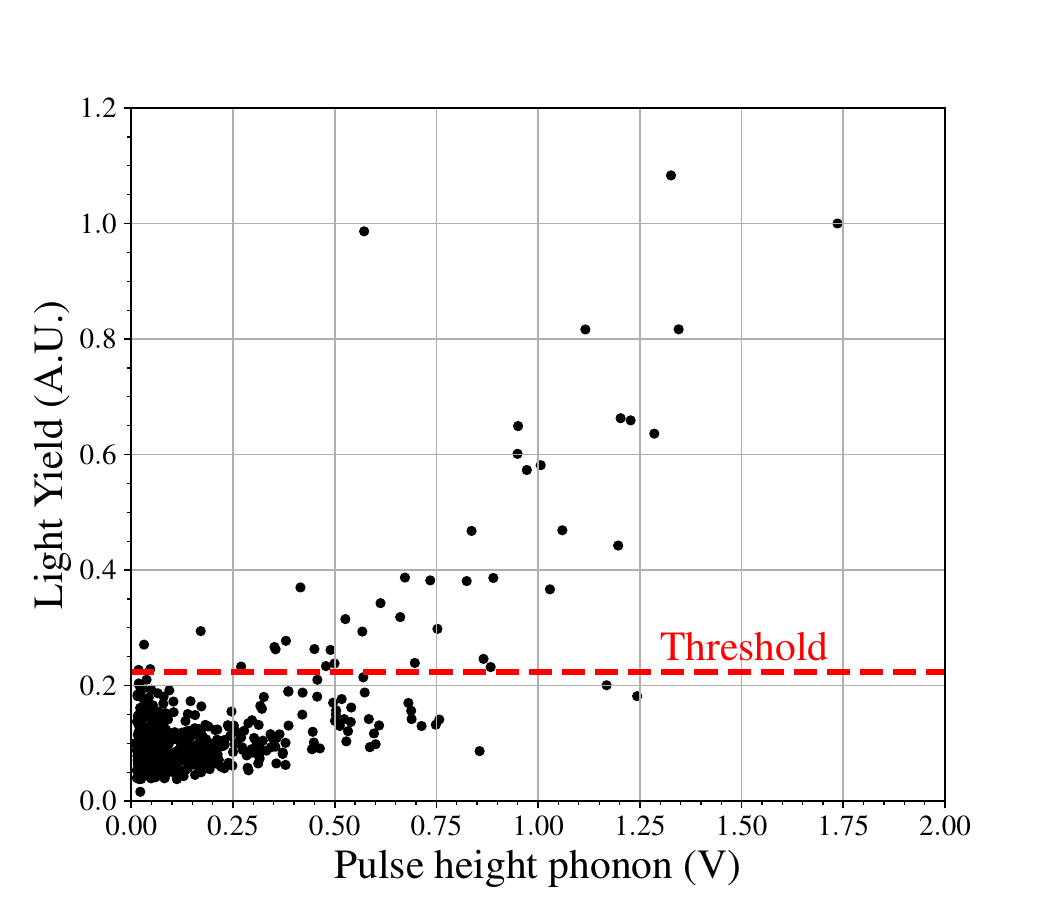}
    \caption{Optimum filter amplitudes of coincident events in the sugar crystal and its light detector. Only the phonon channel was used for triggering. The red dotted line indicates the light detector threshold, defined consistently with that of the sugar crystal, and is shown solely to indicate the light-detector noise level for visual comparison. A clear amplitude correlation emerges events above $\sim$0.5 V in the sugar crystal, indicating light generation from an energy deposition inside the sugar crystal.}
    \label{fig:sugar-LY}
\end{figure}

Detailed inspection of the data shows a significant number of events in which a signal in the light channel occurs within a few milliseconds of the onset of a signal in the phonon channel, an example of which is shown in Figure \ref{fig:sugar-pulse}. These coincidence events are observed exclusively for pulses with amplitudes exceeding $\sim$0.5\,V in the phonon channel, and appear to be correlated at higher amplitudes, as shown in Figure \ref{fig:sugar-LY} for a representative data file. The observation that a significant number of high amplitude pulses in the sugar coincide with events in the light detector, combined with a correlation between the amplitude in the sugar and light detectors, indicates that the sucrose crystal is producing scintillation light for particle interactions of sufficient energy.  These preliminary results motivate further examination of sugar as a particle detector for light dark matter.

\section{Conclusions}

The nature of dark matter remains one of the most intriguing open questions in contemporary particle physics. The detection of light dark matter, with masses below 1 GeV/c$^2$, requires developing detectors sensitive to very low momentum transfers.  The use of targets with light nuclei, such as hydrogen, enhances sensitivity to the lowest masses of dark matter, which is the focus of the SWEET project.


In this work, we investigate the potential of sugar, an organic material rich in hydrogen, as a cryogenic detector in the search for light dark matter.  To assess the feasibility of using sugar as a target material, a monocrystalline sucrose crystal was grown, instrumented with an NTD thermistor and operated in the presence of a light detector. During the measurement campaign, thermal pulses were observed in the sugar crystal, along with a significant number of coincidences between the phonon and light channels.  This suggests a scintillation response in sucrose, a feature which is  relevant for background rejection and event identification in similar cryogenic detectors \cite{kinast2023characterisation}. These results demonstrate that sugar can be operated as a cryogenic calorimeter, and may serve as a promising target material in the direct search of light dark matter.

In the future, larger crystals will be grown using higher-purity sucrose for improved material quality. The detector will be updated by replacing the NTD sensor with a TES to enhance sensitivity. A calibration source will also be introduced to enable precise energy reconstruction and further characterize the detector response. An upgrade to the crystal growth facility is currently underway to support these developments.

The SWEET project has opened a new avenue for the development of cryogenic detectors based on organic crystals, and highlights the potential of sugar-based materials to probe the low-mass dark matter parameter space.


\printbibliography

\end{document}